# Micro-Doppler Frequency Comb Generation by Axially Rotating Scatterers

V. Kozlov[1,a),b)], D. Filonov[1,2 ,b)], Y. Yankelevich[1] and P. Ginzburg[1,2]

[1]School of Electrical Engineering, Tel Aviv University, Tel Aviv, 69978, Israel

[2]ITMO University, St. Petersburg 197101, Russia

**Abstract**:

Electromagnetic scattering in accelerating reference frames inspires a variety of phenomena, requiring employment of general relativity for their description. While the 'quasi-stationary field' analysis could be applied to slowly-accelerating bodies as a first-order approximation, the scattering problem remains fundamentally nonlinear in boundary conditions, giving rise to multiple frequency generation (micro-Doppler shifts). Here a frequency comb, generated by an axially rotating subwavelength (cm-range) wire and split ring resonator (SRR), is analyzed theoretically and observed experimentally by illuminating the system with a 2GHz carrier wave. Highly accurate 'lock in' detection scheme enables factorization of the carrier and observation of more than ten peaks in a comb. The Hallen integral equation is employed for deriving the currents induced on the scatterer at rest and a set of coordinate transformations, connecting laboratory and rotating frames, is applied in order to predict the spectral positions and amplitudes of the frequency comb peaks. Unique spectral signature of micro-Doppler shifts could enable resolving an internal structure of the scatterers and mapping their accelerations in space, which is valuable for a variety of applications spanning from targets identification to stellar radiometry.

---

[a)] vitaliko@mail.tau.ac.il
[b)] V. Kozlov and D. Filonov contributed equally to this work

**Introduction**

Laws of physics in accelerated frames of reference could be substantially different from those obtained in inertial ones. Coriolis force and Sagnac interference are, probably, the most known effects in mechanical and electromagnetic systems correspondingly. Electromagnetic scattering from moving bodies requires special formulation, traced back to investigation of rotating machinery and radar targets identification. Special attention has been paid to studies of non-relativistic rotating objects, owing both to the relative mathematical simplicity of their analysis and to the vast span of potential applications. Among various examples it is worth noting radio frequency (RF) detection of helicopter propellers [1],[2],[3], back action of spinning blades on wireless communication channels [4] and even rotational motion measurement of celestial bodies in astrophysics [5]. Moreover, acoustical noise could originate from revolving motors [6], rotating and vibrating wires were proposed as an analogue to the molecular Raman scattering in optics [7], while, in general, signatures of rotation emerge in many others multidisciplinary fields. In a broader scenes, time-varying scatterers could find use in radio frequency identification (RFID) technologies [8],[9].

Rigorous analysis of electromagnetic scattering from moving and, in particular, accelerating bodies requires employment of apparatus developed for the general theory of relativity [10]. Along with the transformation of all physical observables to the moving frame, accelerated motion causes substantial change in the general form of the governing differential equations and their associated boundary conditions. A relativistic treatment of electromagnetic fields in the presence of rotating bodies (infinite two dimensional cylinders) was performed by Bladel [10] and few others. Intuitively, each point on a rotating scatterer has a different instantaneous velocity, resulting in a collection of Doppler shifts. As electromagnetic fields at nearby points on a body are tightly connected by the nonlocal nature of Maxwell's equations, the scattering problem is far from being trivial. Furthermore, time-varying boundary conditions make the electromagnetic problem nonlinear, giving rise to generation of new frequencies alongside the initial carrier wave spectrum. While solution of the general scattering problem is quite involved, a set of simplifications could be made by assuming nonrelativistic motions and slow accelerations. First order approximations are widely used in theoretical analysis of radar targets identification, e.g. assuming that objects are at a complete rest during the interaction with an incident pulse [11]. Collecting static scattered signals, obtained from a few spatial positions of a target, enables relating theoretical data to experimental - radar measurements usually provide phase relations between adjutant pulses impinging on a target, or Doppler shifts in the reflected wave spectrum. Accelerating bodies, in contrast, produce a much richer spectral signature, called micro-Doppler shifts [12], that cannot always be recovered by applying the static approximation. The so-called 'instantaneous rest frame' technique assumes slow nonrelativistic motion as well, but relaxes few of the beforehand mentioned constraints and enables extracting more information regarding the accelerated motion. Neglecting second order relativistic contributions enables factoring out time dilation and length contraction effects and employing standard field continuity boundary conditions at each point of a scatterer, as if it moves with a linear (instantaneous) velocity. The benefit of rotating systems is in the existence of a relatively simple frame, where the entire scatterer is at rest. In this case the route to undertake is to transform sources to the rest frame of the rotating body, solve the scattering problem, and translate the results back to the laboratory frame. This approach is undertaken here and compared to the experimental data.

While much work was put into the description of various time varying scatterers (see for example [13],[14],[15],[16]), the dipole approximation for scattering from objects that are smaller than the incident wavelength is both simple and useful. Here the effect of electromagnetic scattering from rotating objects (electric - thin wire, and magnetic - SRR) is studied both experimentally and analytically, including retardation effects that arise due to the small, yet finite, dimensions of the scatterers. To the best of our knowledge a theoretical treatment of scattering from a clutter of random rotating dipoles was partially addressed in [17]. In particular, a dipole, illuminated by a field of frequency ω was considered, predicting the scattered field to have additional components of $\omega \pm 2\dot{\theta}$, with $\dot{\theta}$ being the angular frequency of rotation. In a later work a similar result was reached theoretically as well [18]. While the major Doppler shift indeed occurs at $\pm 2\dot{\theta}$ frequencies, rotating bodies with nontrivial internal structure generate an entire frequency comb of micro-Doppler shifts. Detailed analysis, presented here, shows the emergence of the frequency comb in the scattered field, having components of $\omega \pm m\dot{\theta}$, where m is an integer. A nonrelativistic model, based on Hallen's integral equation [19] for calculating the current distribution on thin conducting wires, was employed together with the 'instantaneous rest frame' approximation, under appropriate coordinate transformations between lab and scatterer's reference frames. The theoretical model was fully supported by the experiment undertaken at an anechoic chamber, where shaped copper wires were illuminated with 2GHz waves and the micro-Doppler comb was measured with highly accurate 'lock in' detection scheme.

**Results**

*Theory*

Theoretical analysis of a spinning wire in free space will be presented first and then verified experimentally. Demonstration of a rotating SRR, a magnetic analogue of the electric dipole, will follow. The geometry of a spinning scatterer in free space and relevant measurement apparatus appear in Fig. 1. For the theoretical analysis, consider a short wire, rotating in the Z-Y plane at a constant angular frequency $\dot{\theta} \ll \omega$, where $\omega$ is the carrier frequency of the incident illumination. In the lab reference frame (denoted by prime) an incident plane wave of the form $E^{inc} = E_0 e^{ikz'} \hat{y}'$ is propagating in free space and the carrier's harmonic time dependence $e^{i\omega t}$ is suppressed. The rest frame of the wire (unprimed) agrees with the lab coordinates at time $t = 0$. The primary goal of the derivation is to calculate time-dependent forward scattering. In the considered geometrical arrangement the projection of the wire on the polarization direction changes harmonically in time (from maximal to almost zero), resulting in time-dependent scattering.

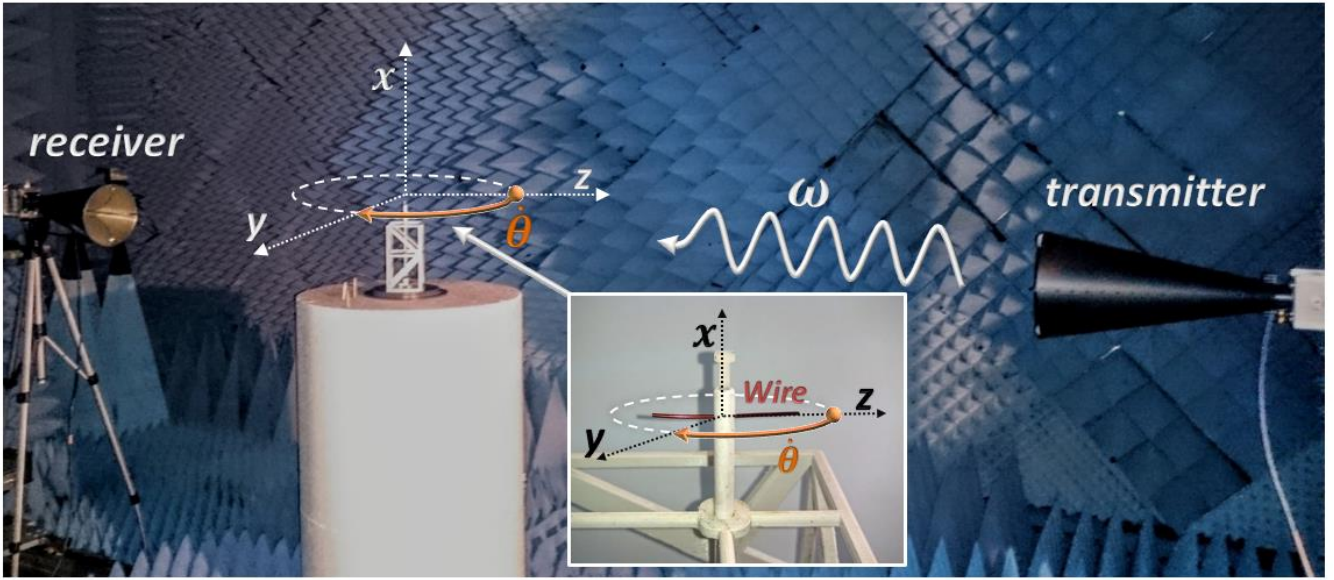

Figure 1: Experimental setup of transmission in an anechoic chamber. A scatterer is mounted on a motor's shaft and placed between transmitting and receiving horns, located in the far field of the scatterer.

The coordinate transformation between the lab and the rotating frame is carried out by applying the following relation:

$$\begin{pmatrix} x \\ y \\ z \end{pmatrix} = \begin{pmatrix} 1 & 0 & 0 \\ 0 & Cos(\dot{\theta}t) & -Sin(\dot{\theta}t) \\ 0 & Sin(\dot{\theta}t) & Cos(\dot{\theta}t) \end{pmatrix} \begin{pmatrix} x' \\ y' \\ z' \end{pmatrix}. \tag{1}$$

In the wire's rest frame the incident field along the length of the wire becomes $E_z^{inc}(z,t) = E_0 Sin(\dot{\theta}t) e^{ikCos(\dot{\theta}t)}$ (the wire is parallel to the z-axis in its frame of reference, hence only z-component of the field is relevant for the interaction). The field in the scatterer's rest frame may be decomposed into a sum of products of Bessel functions, dependent on space-time harmonics:

$$E_z^{inc}(z,t) = E_0 Sin(\dot{\theta}t) \sum_{n=-\infty}^{\infty} J_n(kz) e^{in(\dot{\theta}t - \frac{\pi}{2})}. \tag{2}$$

Scattering problems from thin perfectly conducting wires could be solved by applying Hallen's integral formulation. Scatterers are virtually replaced by line currents (unknowns) and the total electromagnetic field outside is given as a superposition of the excitation and the convolution between the current and the Green's function related kernel. Boundary conditions of vanishing tangential components of the electrical field on a perfect electric conductor (physical boundaries of the initial wire) enable formulating the scattering problem in the form of an integral relation. Hallen's integral equation under the assumption of a lossless, perfectly electrically conducting wire is approximated by [19]:

$$\tilde{Z} I(z,t) = C_1 e^{ikz} + C_2 e^{-ikz} + i \int_{-h}^{h} e^{-ik|z-z'|} E_z^{inc}(z',t) dz', \tag{3}$$

where $\tilde{Z} \approx \frac{i\eta e^{-ika}}{2\pi a}$, $a$ is the radius of the wire, $\eta$ is the free space impedance and $C_{1,2}$ are to be determined from the boundary conditions of vanishing current at the edges of the wire, located at $z = \pm h$. Solving equation (3) enables determining the current along the wire:

$$I(z,t) = \frac{4}{\tilde{Z}k} \sum_{n=0}^{\infty} T_n^c(\dot{\theta}t)\left(S_n^c(kz) - S_n^c(kh)\frac{Cos(kz)}{Cos(kh)}\right) + iT_n^s(\dot{\theta}t)\left(S_n^s(kz) - S_n^s(kh)\frac{Sin(kz)}{Sin(kh)}\right), \tag{4}$$

where the following auxiliary functions are defined as:

$$\begin{aligned}
T_n^c(\dot{\theta}t) &= E_0 \frac{(-1)^n(2-\delta_{n,0})}{(2n+1)!} Sin(\dot{\theta}t)Cos(2n\dot{\theta}t), \\
T_n^s(\dot{\theta}t) &= 2E_0 \frac{(-1)^n}{(2n+2)!} Sin(\dot{\theta}t)Cos((2n+1)\dot{\theta}t), \\
S_n^c(kz) &= \left(\frac{kz}{2}\right)^{2n+1} \left(F_R^{(2n)}(kz)Sin(kz) - F_I^{(2n)}(kz)Cos(kz)\right), \\
S_n^s(kz) &= \left(\frac{kz}{2}\right)^{2n+2} \left(F_R^{(2n+1)}(kz)Sin(kz) - F_I^{(2n+1)}(kz)Cos(kz)\right),
\end{aligned} \tag{5}$$

while $F_{R/I}^{(m)}$ are the real and imaginary components of the hypergeometric function [20], satisfying:

$$F_R^{(m)} + iF_I^{(m)} = {}_2F_2(\{m+0.5, m+1\}, \{m+2, 2m+1\}, 2ikz). \tag{6}$$

Equations (4) and (5) show that the current consists of both even and odd time harmonic components of $\dot{\theta}$ that produce additional scattered field frequencies in the far field. Next the scattered field in the rest frame of the wire is derived and transformed back into the lab frame of reference using equation (1), evaluating it in the lab frame at a point $z' = -d, y' = 0, x' = 0$ (where the detector is located) and taking the $\hat{y}'$ component alone (as is the case in our experiment):

$$E_{y'}^{Lab}(d,t) = \mu \hat{y}' \cdot \int_{-h}^{h} \overleftrightarrow{G}_{FF}(\mathbf{r};z) I(z,t) \hat{z} dz$$
$$= -\mu \int_{-h}^{h} \frac{e^{-ikR(z,d,t)}}{4\pi R(z,d,t)^3} \left( d^2 Sin(\dot{\theta}t) - \frac{zd}{2} Sin(2\dot{\theta}t) \right) I(z,t) dz, \quad (7)$$

where $R(z,d,t) = \sqrt{d^2 - 2zd Cos(\dot{\theta}t) + z^2}$ is the distance connecting a point on the wire and the detector's location, $\overleftrightarrow{G}_{FF}(\mathbf{r};z)$ is the far field 3D Green's dyadic in the rest frame and $\mu$ is the free space permeability. Equation (7) may be solved numerically, or evaluated analytically by adopting several approximations. Here, leading terms in $\frac{1}{d}$ are considered along with the short wire approximation ($kh \ll \frac{\pi}{2}$) for calculating induced currents, which enables identifying well-defined spatial parities in $\dot{\theta}$ series by making Tailor expansions of equations (5). Here even spatial contributions could be approximated as constants, while the odd ones may be replaced with functions that are linear in their argument. The resulting short wire current then takes the form:

$$I_s(z,t) \approx \frac{4}{\tilde{Z}k} \sum_{n=0}^{\infty} \alpha_n^c T_n^c(\dot{\theta}t) \left( \left(\frac{kz}{2}\right)^{2n+2} - \left(\frac{kh}{2}\right)^{2n+2} \right) + i\alpha_n^s T_n^s(\dot{\theta}t) \left( \left(\frac{kz}{2}\right)^{2n+3} - \frac{z}{h}\left(\frac{kh}{2}\right)^{2n+3} \right), \quad (8)$$

where the following auxiliary functions are defined:

$$\alpha_n^c = 2F_R^{(2n)}(0) - 4\frac{\left(2n+\frac{1}{2}\right)(2n+1)}{(2n+2)(4n+1)} {}_2F_2\left(\left\{2n+\frac{3}{2}, 2n+2\right\}, \{2n+3, 4n+2\}, 0\right),$$
$$\alpha_n^s = 2F_R^{(2n)}(0) - 4\frac{\left(2n+\frac{3}{2}\right)(2n+2)}{(2n+3)(4n+3)} {}_2F_2\left(\left\{2n+\frac{5}{2}, 2n+3\right\}, \{2n+4, 4n+3\}, 0\right). \quad (9)$$

It can be readily checked that for small values of $kh$ Eq. (8) and (4) coincide. By using the above mentioned approximations, the scattered field is finally evaluated in the lab reference frame by solving:

$$E_{y\prime}^{Lab}(d,t) \approx \frac{\mu}{4\pi d}\int_{-h}^{h}Sin(\dot{\theta}t)I_s(z,t)dz,$$

$$= \frac{4\mu E_0}{\pi \tilde{Z}k^2}\frac{e^{-ikd}}{d}\sum_{n=0}^{\infty}C_n\left(\frac{kh}{2}\right)^{2n+3}Sin^2(\dot{\theta}t)Cos(2n\dot{\theta}t), \tag{10}$$

where

$$C_n = \frac{(-1)^n(2-\delta_{n,0})}{(2n+1)!}\frac{2n+2}{2n+3}\alpha_n^c. \tag{11}$$

It is apparent that the analytical expression in (10) only recovers the even harmonics of the scattered field. The odd harmonics vanish due to parity considerations involving odd and even functions in the current along with symmetric integration boundaries of the integral. These harmonics disappear even if the full current equation (4) is used. In fact, the odd harmonics will vanish even if the only approximation made is $R(z,d,t) \approx d - zCos(\dot{\theta}t)$ (quadratic terms in wire's internal geometry are neglected), but the detailed proof of this statement is omitted in this paper in order to not obscure the main results. As an important remark, the above observation indicates that an internal geometry of a scatterer has the direct influence on odd micro-Doppler peaks.

*Experimental Results*

Micro-Doppler shifts in the scattered spectra of thin wires will be investigated experimentally and compared with the developed theory. The photograph of the experimental setup appears in Fig. 1. Spinning scatterers, shown as insets in Fig. 2, were mounted on a motor's shaft, stabilized with a 3D-printed plastic rack. The motor is controlled by a DC current, enabling rotational speeds of up to hundreds of hertz. In particular, a 5cm long wire and a SRR, made of the same type of wire, were rotated in the Z-Y plane at a constant angular frequency $\dot{\theta} = 2\pi \times 18\ Hz$. The whole structure was placed between two horn antennas, separated by a 2m distance, ensuring that both the transmitting and receiving horns are at the far-field region of the scatterer.

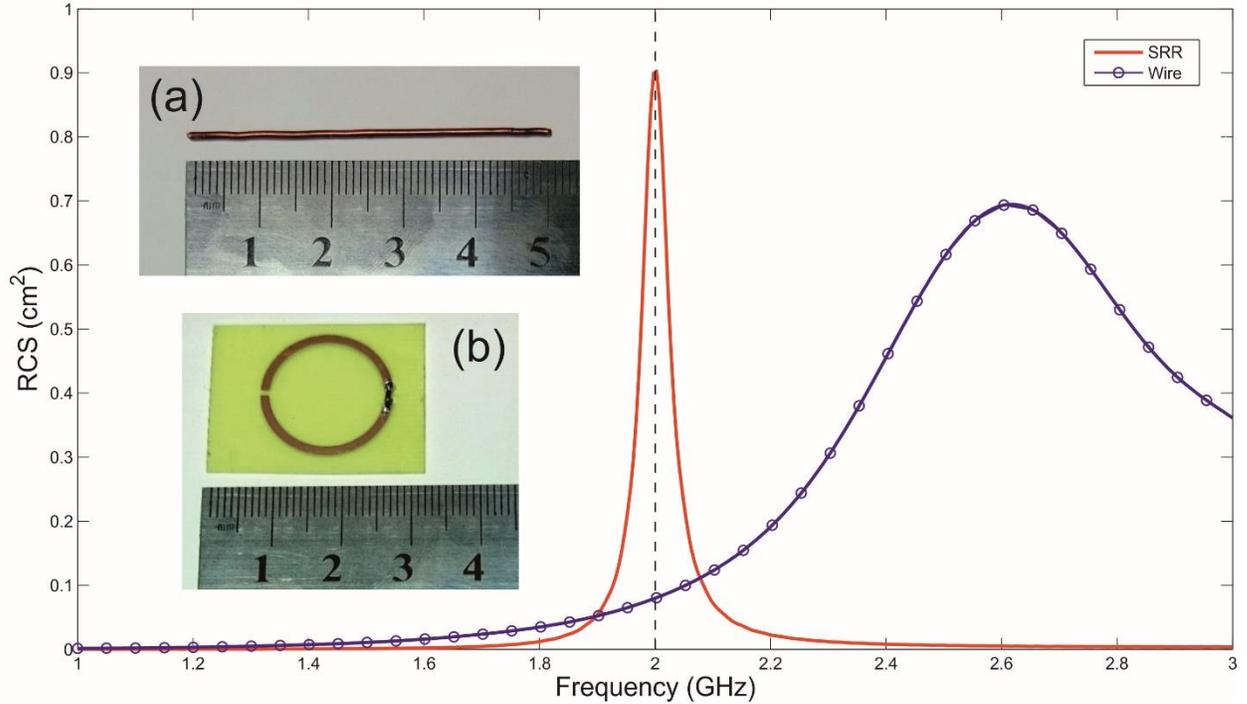

Figure 2: Radar cross sections (numerical simulation) of static scatterers involved in the experiment. (a) wire, (b) split ring resonator.

A Vector Network Analyzer (VNA) was used as a source of a 2GHz continuous wave, feeding the transmitting antenna with the polarization set in the Y-direction. A receiving antenna was placed opposite the transmitter with the same polarization. The output of the receiver was amplified and down-converted to the baseband by mixing it with the same 2 GHz carrier, as in the incident field. The output of the mixer was passed through a low pass filter to remove residual high frequencies and directed to a lock-in amplifier where the low-frequency spectral shifts of the scattered field were recorded. The results are shown in Fig. 3, showing normalized lock-in readouts as a function of baseband frequency. The amplitude of the second peak was used to scale the theoretic values. This peak corresponds to $\omega \pm 2\dot{\theta}$ in the original signal and appears as $2\dot{\theta}$ in the lock-in detection, since only the positive part of the spectrum is presented.

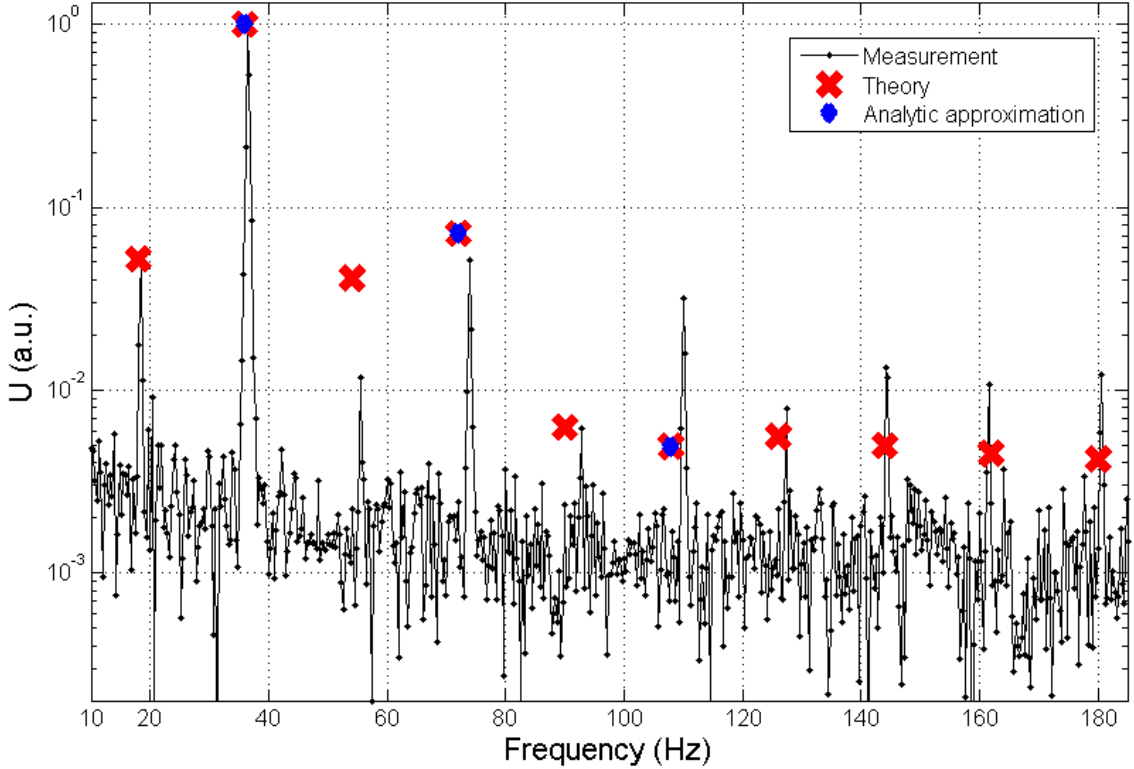

Figure 3: Normalized forward scattering spectrum (micro-Doppler comb, baseband spectrum) of a 5cm long wire, rotating at an angular frequency $\dot{\theta} = 2\pi \times 18\ Hz$. Black line with dots represents the experimental data, red crosses – theoretical predictions evaluated numerically (Eq. 7), blue rhombuses – analytical approximation of even peaks (Eq. 10).

Black-doted curves, representing the experimental data in log scale, show well defined peaks at $m\dot{\theta}$ frequencies, with intensities above the noise level. Ten micro-Doppler peaks were found and others in the comb could be detected by increasing the frequency range and acquisition time of the experiment. The set of red crosses corresponds to the theoretical prediction, obtained with Eq. (7), while blue rhombuses account for the simplified analytic approximation for the even peaks (Eq. 10). From Fig.3 it is apparent that the theory (up to inherent limitations) agrees well with the location and amplitudes of the additional rotational Doppler shifts in the scattered field spectra. Spectral deviations are attributed to fluctuations in the frequency of rotation, which occurred due to the long measurement time (up to a few minutes for each data point was needed in order to obtain higher signal to noise ratios) as well as related instabilities of the motor's frequency of rotation. The amplitude of the peaks is in good agreement with the numerical solution of Eq. (7). The analytic approximation of Eq. (10) is close for the first three even peaks but becomes much smaller for the 4th and 5th even peaks, suggesting that higher orders are needed in the approximations leading to Eq. (10).

A split ring resonator (SRR) is the magnetic counterpart of an electric dipole. Circular current in the ring creates magnetic flux, while the split enables tuning the resonant behavior of the whole structure. As a result, this subwavelength magnetic dipole could have a substantial radio cross section (RCS), making it an attractive building block for RF metamaterials applications (e.g. [21], [22]). Exactly the same experiment, as was performed in the case of the subwavelength wire, was repeated for a SRR, tuned to resonate at the carrier frequency of 2GHz by using a varactor diode that enabled fine tuning the resonant frequency. The photograph and relevant geometrical dimensions appear in Fig. 2 (inset b). The recorded baseband spectrum is shown in Fig. 4.

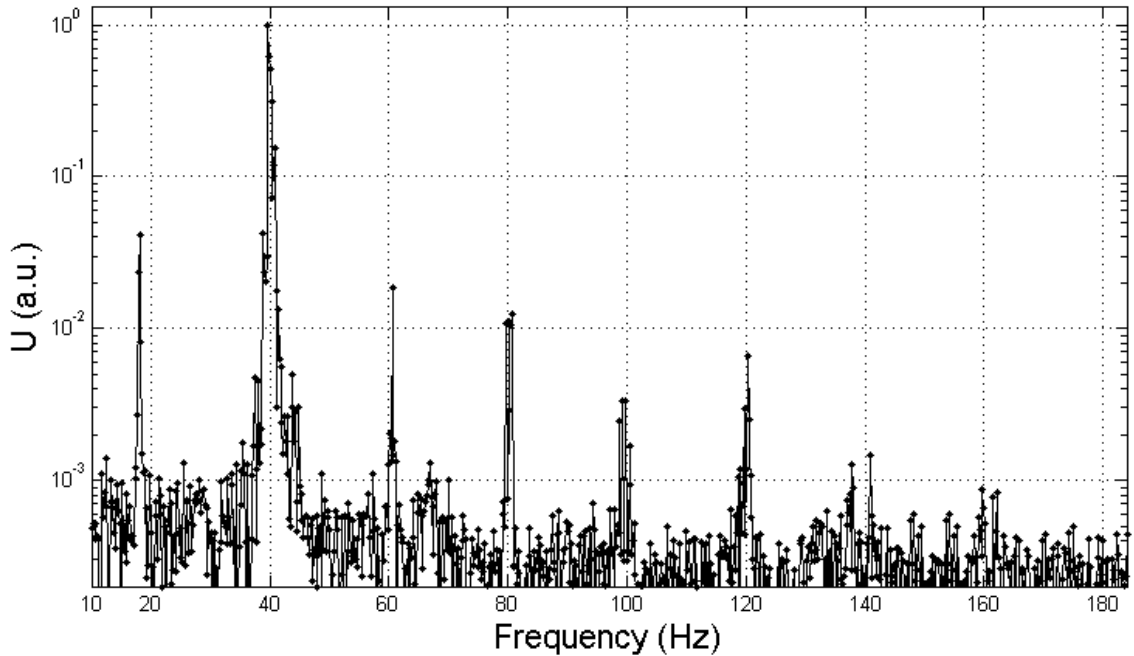

Figure 4: Normalized forward scattering spectrum (micro-Doppler comb, baseband spectrum) of the scattered field from a SRR tuned to resonate at 2GHz, rotating at an angular frequency $\dot{\theta} = 2\pi \times 18\ Hz$.

The scattered baseband spectrum from a SRR has remarkable similarity to the one of the short wire, shown in Fig. 3. Both exhibit discrete scattered frequencies $\omega \pm m\dot{\theta}$ with a similar amplitude distribution, decaying at the higher frequencies of the comb. Hallen's integral formulation enables deriving a set of formulas, similar to those describing a wire. This task, however, is outside the scope of this report owing to the complexity of the involved geometry. The duality principle of Maxwell's equations (e.g. [19]) suggests similarities in the response of electric and magnetic dipoles to an incident field. Note that while the electric dipole should be parallel to the electric field in order to get the strongest interaction, it's the normal to the SRR that should be parallel to the magnetic component of the incident plane wave. While the duality principle is strictly correct for dipoles, the scatterers considered in this paper have finite dimensions, which account for their different spectral characteristics. This difference suggests that it might be possible to use such spectral signatures for resolution of the internal structure of the observed scatterer. Additional strict difference between the wire and SRR, considered here, is their RCS at the carrier frequency of 2 GHz. While the first object is far from its resonance, the second one was tuned to exhibit it exactly at 2GHz (see Fig. 2, finite elements numerical simulation was performed with CST Microwave Studio). As a result, the field scattered from the SRR, detected at the lock-in, is much stronger than in the case of the wire, significantly reducing the acquisition time required for obtaining reliable data.

**Outlook and Conclusion**

In conclusion, it was shown experimentally that the scattered electromagnetic spectra from axially rotating bodies without a symmetry of revolution (thin wire and SRR were considered here) have the frequency components of $\omega \pm m\dot{\theta}$, forming a micro-Doppler frequency comb (m is any integer). A theoretical model for this phenomenon in the case of thin short wires was derived, predicting the appearance of the scattered frequency components and closely predicting their relative magnitudes. These results expand previously proposed theories, where only two Doppler shifts of $\omega \pm 2\dot{\theta}$ were expected in the spectra of dipole-like scatterers, which suggests that it might be possible to use such spectral signatures for resolution of the internal structure of observed scatterers.

**Acknowledgments:**

The Authors acknowledge Dr. G. Slepyan (Tel Aviv University) for his valuable advice on theoretical aspects and Yossi Kamir (Tel Aviv University) for 3D printing parts of the experimental setup.

This work was supported, in part, by TAU Rector Grant and German-Israeli Foundation (GIF, grant number 2399). This work has been partially supported by Government of the Russian Federation (Grant No. 074-U01). 'Quasi-stationary field' analysis of scatterers' mechanical motion has been partially funded by the Russian Science Foundation (Grant No. 14-12-01227).